\documentclass[aps,pra,twocolumn,showpacs,amsmath,amssymb]{revtex4}
\usepackage{graphicx}
\usepackage{dcolumn}
\usepackage{bm}
\begin{document}

\title{Laser tweezers for atomic solitons.}
\author{Alicia V. Carpentier$^{1}$, Juan Belmonte-Beitia$^{2}$, 
Humberto Michinel$^{1}$ and V\'{\i}ctor M. P\'erez-Garc\'{\i}a$^{2}$}
\affiliation{$^{1}$\'Area de \'Optica, Facultade de Ciencias de Ourense,\\ 
Universidade de Vigo, As Lagoas s/n, Ourense, ES-32004 Spain.}
\affiliation{$^{2}$Departamento de Matem\'aticas, E. T. S. I. Industriales,\\ and Instituto de Matem\'atica 
Aplicada a la Ciencia y la Ingenier\'{\i}a (IMACI),
Universidad de Castilla-La Mancha, 13071 Ciudad Real, Spain.}

\begin{abstract}
We describe a controllable and precise laser tweezers for Bose-Einstein condensates of ultracold
atomic gases. In our configuration, a laser beam is used to locally modify the sign of the scattering
length in the vicinity of a trapped BEC. The induced attractive interactions between atoms allow to 
extract and transport a controllable number of atoms. We analyze, through numerical simulations, 
the number of emitted atoms as a function of the width and intensity of the 
outcoupling beam. We also study different configurations of our system, as the use of moving beams. 
The main advantage of using the control laser beam to modify the nonlinear interactions in comparison 
to the usual way of inducing optical forces, i.e. through linear trapping potentials, is to improve the
controllability of the outcoupled solitary wave-packet, which opens new possibilities for engineering  
macroscopic quantum states.

\end{abstract}


\maketitle
\section{Introduction} 
The achievement of Bose-Einstein condensation (BEC) in dilute gases of alkali atoms \cite{1erCBE}
has driven the  research on the design of new tools for the manipulation and
coherent control of atomic ensembles. In the last years, an intensive study of different mechanisms 
for this purpose has been carried out both theoretically and experimentally. Among the most important
contributions we must cite the realization of atom mirrors \cite{mirror}, guides \cite{guide1,guide2},
the design of atomic accelerators both in linear and circular 
geometries \cite{acceleration1,acceleration2,acceleration3}. Atom lasers have also been 
developed, first based on the use of short radio-frequency pulses as an outcoupling 
mechanism, flipping the spins of some of the atoms to release them from the 
trap \cite{laser1}. Later, other coherent atomic sources were built leading to pulsed, 
semicontinuous or single-atom lasers \cite{laser2,laser3,laser4,laser5,laser6,laser7}.

The control of coherent atomic beams is a challenging problem in physics due to its 
potential applications in multiple fields like atom interferometry \cite{interferometry}, 
superposition of quantum states \cite{control}, atom clocks \cite{clock}, or quantum 
information \cite{quantum}, among others. Many of these devices take advantage
of nonlinear interactions between atoms, which are ruled by the value of the scattering length $a$.
The adequate tunning of this parameter has been possible with the use of Feshbach resonances \cite{feshbach} 
and has yield impressive effects like the macroscopic collapse of matter waves \cite{bosenova} or
the creation of atomic solitons \cite{solitons,Litium,ENS}. The recent realization of optical
control of Feshbach resonances \cite{optical} has paved the way for the experimental demonstration 
of many theoretical proposals on nonlinear waves in Bose-Einstein condensates with spatially 
inhomogeneous interactions \cite{tuti,PGPG} including the dynamics of solitons when the modulation of 
the nonlinearity is a random 
\cite{Garnier}, linear   \cite{Panos}, periodic \cite{Boris2},  localized function \cite{Primatarowa} 
or step-like function \cite{Egor}.

In this paper, we propose the use of spatially-dependent scattering length as a tool for 
designing precise atom tweezers which are able to extract a portion of atoms 
from a BEC. Other methods have been proposed for this purpose, as in Ref. \cite{opticaltweezers} 
and Ref. \cite{modulator} in combination with spatial light modulators for splitting the atomic cloud. 
Our device is inspired on a coherent atomic source based on the spatial modification 
of the scattering length \cite{laser8,laser9}, which produces a highly regular and controllable 
number of atomic pulses by modulating $a$ along the trapping axis of a BEC. As we will show 
in this paper, in comparison with linear traps which do not alter the value of $a$, 
spatially-tailored nonlinear interactions yield robust control of the number of atoms that 
can be extracted from a BEC reservoir, providing new ways to the creation of macroscopic 
superposition of quantum states from arrays of Bose-Einstein condensates \cite{phillips}.

The structure of this paper is as follows: in Section \ref{section2}, we introduce the configuration of 
the system and the mean field theory which is used in this work. In Section \ref{section3} we study by 
means  of numerical simulations the mechanism of emission by employing a static laser tweezers. We also 
analyze the number of atoms emitted as a function of the main parameters of the system. Finally, in Section 
\ref{section4},  we analyze the use of moving tweezers which trap atoms by crossing the condensate, and compare 
our results with the linear case in which there is no spatial variation of the scattering length.
 
\begin{figure}[htb]
{\centering \resizebox*{1\columnwidth}{!}{\includegraphics{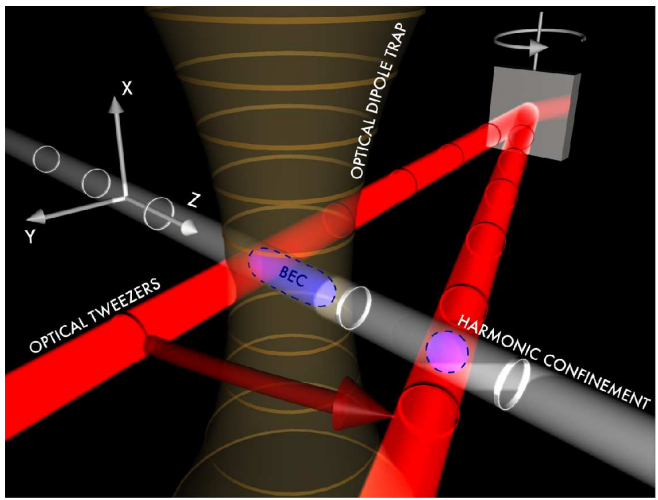}} \par}
\caption{[Color online]. Sketch of the system we will study in this
paper for the case of optical tweezers which are used to extract
a given number of atoms from a BEC reservoir. A rotating mirror is
used to move a laser beam over a BEC which is transversely trapped 
by magnetic confinement, and optically along the direction of motion of the laser.}
\label{fig1}
\end{figure}


\section{System studied and model equations}
\label{section2}
We will assume that a BEC is strongly trapped in the transverse directions $(x,y)$ and
weakly confined along the longitudinal one $(z)$ leading to a {\em cigar shaped} configuration.
We will consider the effect of a spatial variation of the scattering length along $z$ which can
be switched from positive to negative values by the optical control of Feshbach resonances by 
means of a laser beam. The region of negative 
scattering length can be varied in size and displaced along $z$ by simply focusing or moving the laser beam. 
The choice of an optical control \cite{optical} instead of a magnetic one \cite{magnetic} allows 
for a faster and easier manipulation of the spatial variations of the scattering length.

The mean field description of the dynamics of the BEC is provided by a Gross-Pitaevskii (GP) 
equation of the form:
\begin{equation}
\label{GP}
i\hbar\frac{\partial\Psi}{\partial t}=-\frac{\hbar^2}{2m}\Delta^2\Psi+V(\vec{r})\Psi+U|\Psi|^2\Psi,
\end{equation}
where $\Psi$ is the order parameter, normalized to the number $N$ of atoms in the cloud: 
$N=\int|\Psi|^2d^3r$. $U=4\pi\hbar^2a/m$ characterizes the 2-body interactions determined by the value 
of the scattering length $a$. The cloud of $N$ equal bosons of mass $m$ is tightly trapped in $(x,y)$ 
by a harmonic potential $V_\perp$ of frequency $\nu_\perp$ and weakly confined along $z$ by the effect 
of an optical dipole trap $V_z$ that can be produced by a laser beam of a given width 
along $z$ \cite{pot1,pot2}. The mathematical expression for the potential is thus:
\begin{multline}
\label{pot}
V(\vec{r}) = V_\perp+V_z= \nonumber  \\
=\frac{m\nu^2_\perp}{2} \left( x^2+y^2 \right)+V_d
\left[1-\exp\left(-\frac{z^2}{L^2} \right)\right],
\end{multline}
where $V_d$ is the depth of the (shallow) optical dipole potential and $L$ its characteristic width 
along $z$. To fix ideas, we will present specific numbers in this paper corresponding to $^7$Li, using
the experimental parameters of Ref. \cite{Litium}, where $V_d\approx\hbar\nu_\perp$, being $\nu_\perp=1$ kHz the 
frequency of the trap in the transverse plane ($x,y$), which yields to a transverse radius $r_\perp\approx 3  \mu$m. The 
other numerical values used in our simulations are $L=4r_\perp$, $N=3\cdot 10^5$, $w_c=5.4r_\perp$ 
and $a=-1.4$ nm. We must stress that our basic ideas should hold for different atomic species like $^{85}$Rb 
and $^{133}$Cs with an adequate change of the parameters used.

\section{Atom extraction with static tweezers and control of the atomic wavepacket}
\label{section3}
The problem we will face is the  controllable extraction of atoms from a BEC
reservoir. We will consider a system configuration in which a trapped cigar-type condensate is partially 
overlapped by a laser beam in a similar configuration to the one described in Ref. \cite{laser8}. Under adequate conditions, 
the laser changes the local value of the scattering length in part of the cloud. If $a$ is locally switched 
to large enough negative values, a burst of atomic solitons can be emitted from the condensate. If the distance 
between the laser beam and the center of the cloud is kept constant, a emitted soliton may rebound inside the 
laser beam and thus remain trapped out of the reservoir \cite{laser9}. Once the atoms are extracted, the laser 
can be moved away from the condensate. This also allows to control the position of the soliton in the $z$ direction. 
This idea of extracting atoms 
 is radically different than using a usual dipole trap to extract atoms without 
switching the scattering length to negative values. In the later case, the atoms will perform Josephson oscillations 
between the reservoir and the tweezers and it is only possible to extract a significant portion only at times 
exactly matching the maxima of the periodic motion. Thus, the role of nonlinear interactions is essential in this 
static configuration to guarantee that once extracted the atoms will not go back to the reservoir. 

\begin{figure}[htb]
{\centering \resizebox*{1\columnwidth}{!}{\includegraphics{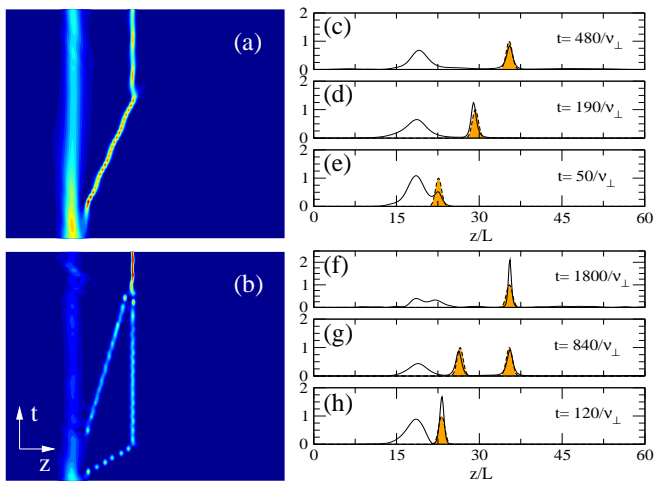}} \par}
\caption{[Color online] Controlled emission of atomic solitons from a BEC reservoir.
The extraction was made by employing a gaussian-profile laser beam. Once the soliton is emitted,
the beam is separated from the condensate dragging the emitted atoms. In (a) one laser beam 
extracts one soliton and controls its position in $z$. In (b) two solitons are extracted with 
two different beams and their paths joined. The range of times, shown in the vertical axis is from $t=0$ to $t=500/\nu_\perp$ 
in (a) and to $t=1800/\nu_\perp$ in (b), where $\nu_\perp=1
$ kHz is the radial trapping frequency. 
The horizontal axis is $60$ times the width $L$ of the optical dipole trap that confines the 
condensate in the $z$ direction. The figures at the right display the profiles of the condensates 
and emitted solitons (in black continuous lines), and of the laser beams (in dotted black lines 
and shaded) for different times.}
\label{fig2}
\end{figure}

Our results are based on numerical simulations of Eq. (\ref{GP}). All results to be presented in 
what follows have been obtained using second order in time split-step pseudospectral solvers, with the 
spatial derivatives being evaluated by Fourier methods \cite{PGLiu}. 
In Figs. \ref{fig2}(a) and \ref{fig2}(b) we show some numerical simulations showing how our nonlinear tweezers work. In both 
pictures, we have plotted pseudocolor images of the cloud density. The horizontal axis is $z$ and vertical 
axis is time. The extraction of atoms is made with a gaussian-profile laser beam. If the scattering 
length is switched to large-enough negative values, part of the cloud is extracted. Once the atoms have left 
the reservoir, the laser beam is moved away from the condensate dragging part of the atoms. In 
Fig. \ref{fig2}(a) the laser beam extracts one soliton and controls its position along $z$. In 
Fig. \ref{fig2}(b) two solitons are extracted with two different laser beams and their paths joined 
at a given point. Time in the vertical axis goes from $t=0$ to $t=500/\nu_\perp$ in Fig. \ref{fig2} (a) and 
from  $t=0$ to $t=1800/\nu_\perp$ in Fig. \ref{fig2}(b), where $\nu_\perp$ = 1 kHz is the radial trapping 
frequency. The horizontal axis spans $60$ times the width $L$ of the optical dipole trap that confines 
the condensate in the $z$ direction. The figures on the right represent the atom density showing the 
profiles of the reservoir and the emitted solitons (in black continuous lines), and of the laser beams 
(in dotted black lines and shaded) for three different times. As it can be appreciated 
our control method allows a robust control on the extracted atoms.

\begin{figure}[htb]
{\centering \resizebox*{1\columnwidth}{!}{\includegraphics{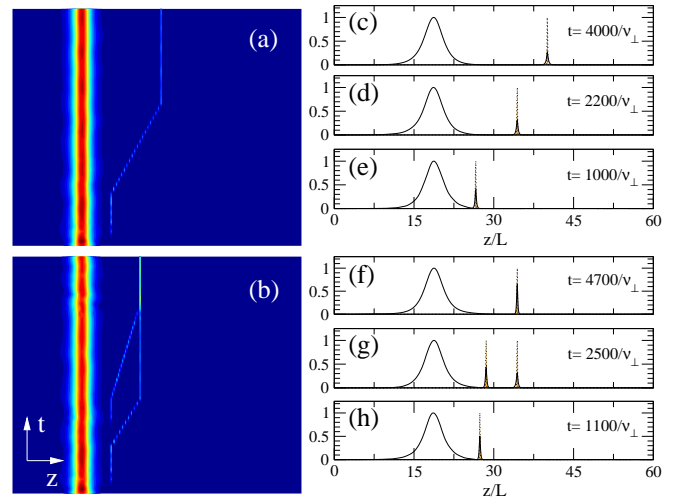}} \par}
\caption{[Color online] Same as in Fig.\ref{fig1} for a narrower and more intense
laser beam. Vertical axis corresponds to time spanning the interval  from $t=0$ to $t=5000/\nu_\perp$. The other parameters are
as in Fig. \ref{fig2}. The figures at the right display the profiles of the condensates and emitted solitons (in black 
continuous lines), and of the laser beams (in dotted black lines and shaded) at three different times 
of propagation. }
\label{fig3}
\end{figure}

Let us now consider a slightly different configuration in order to show the robustness of the method. 
We now address the case of a laser beam which is more intense and narrow 
than in the previous simulations. In this case a high nonlinear interaction stripe is 
generated and some atoms will be attracted to this thin region of negative scattering length 
and will be trapped. As in the previous case, the position of the extracted atoms along $z$ can be controlled by moving the laser. 
In Fig. \ref{fig3}(a) and Fig. \ref{fig3}(b) we show a similar representation as in Fig. \ref{fig2}(a) and 
Fig. \ref{fig2}(b). In this case, a much narrower beam has been used in order to suppress internal
rebounds of the extracted atomic beam. In Fig. \ref{fig3}(b) two matter waves are emitted 
employing two different laser beams. By displacing the lasers, it is possible to control the 
relative position of the extracted solitons. Vertical axis corresponds to times in the range from $t=0$ 
to $t=5000/\nu_\perp$ for both figures. 
The plots at the right display the profiles of the condensate and emitted solitons (in black continuous 
lines) and of the laser beams (in dotted black lines and shaded) for three different times.

\begin{figure}[htb]
{\centering \resizebox*{1\columnwidth}{!}{\includegraphics{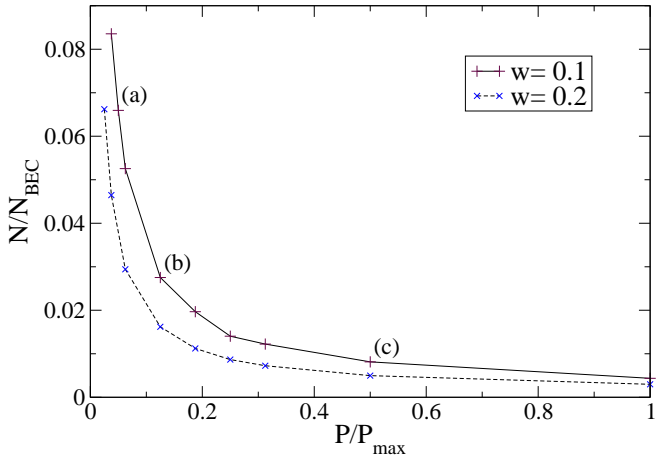}} \par}
\caption{[Color online] Dependence of the percentage of extracted atoms ($N/N_{BEC}$) on the
intensity of the laser beam power ($P/P_{max}$) for two different widths of the outcoupling beam:
$w\approx w_c/50$, plotted with the symbol $+$ (in maroon) and $w\approx w_c/25$, plotted with the symbol 
$\times$ (in blue). The rest of the parameters are indicated in  Fig. \ref{fig5}.}
\label{fig4}
\end{figure}

The number of extracted atoms depends on the laser intensity. In Fig. \ref{fig4} we can see
the percentage of atoms attracted and trapped by the laser tweezers as a function of the intensity
of the laser for two beams of different widths $w$ (much thinner than the longitudinal size of the BEC cloud $w_c$). 
The beam with $w\approx w_c/50$ (plotted with the symbol $+$ in maroon) is able to extract more atoms 
than the beam with $w\approx w_c/25$ (plotted with the symbol $\times$ in blue). As it can be appreciated 
in the plot, the number of atoms extracted decreases as the laser intensity is increased. This is due 
to the fact that the number of emitted solitons increases with the intensity of the laser, thus the 
number of atoms per soliton diminishes.

\begin{figure}[htb]
{\centering \resizebox*{1\columnwidth}{!}{\includegraphics{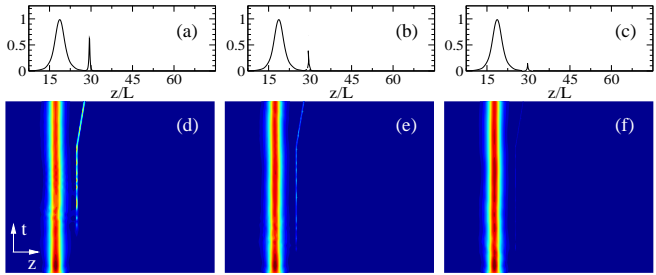}} \par}
\caption{[Color online] (a-c) Examples of atom extraction from a condensate corresponding to different powers corresponding 
to the points labeled with (a), (b) and (c) 
in Fig.\ref{fig4}. Shown are the density profiles for a time propagation of $t=1500/\nu_\perp$.
The laser is separated from its original position at $t=1100/\nu_\perp$.
(d-e) Pseudocolor plot indicating the full evolution of the BEC in the time span $t \in [0, 1500/\nu_{\perp}]$.}
\label{fig5}
\end{figure}

In Fig. \ref{fig5}, we show three different numerical simulations employing beams of different 
intensities. Each figure corresponds with the three points labeled in Fig. \ref{fig4} as (a), 
(b) and (c). The vertical range corresponds to times from $t=0$ to $t=1500/\nu_\perp$. For $t=1100/\nu_\perp$ the 
laser is set into motion along the $z$ axis. The upper profiles show the condensate reservoir and the extracted atoms distributions.

\section{Atom extraction with moving tweezers}
\label{section4}
Another interesting possibility is to use a configuration of moving lasers with variable
velocities. In this case, the extraction of atoms is obtained when the laser traverses the 
condensate, trapping particles through its path. The number of extracted atoms $N_e$ varies with the velocity, 
and with the main parameters of the beam. We have analyzed, by means of numerical simulations, the dependence
of $N_e$ on the beam intensity for several velocities in two different cases: the first one corresponding to 
a moving linear trapping potential and the second one corresponding to our nonlinear tweezers. Physically, the 
first situation consist of an optical dipole trap which does not change the value of the scattering length. 
We assume that in that case the laser which creates this linear trap has the same width and depth as in the 
nonlinear case. The only difference is that nonlinear interactions are suppressed. Our purpose with this comparison
is to evaluate the effect of nonlinear interactions in the extraction procedure. To this aim, we have employed 
the same Schr\"odinger equation  model as in the previous section. In all the simulations, a
laser beam of width $w$ is displaced from $z\ll 0$ to $z\gg 0$ at a fixed given velocity $v$, extracting
a fraction of atoms $N_e/N_{BEC}$ from the reservoir. The BEC reservoir is centered around $z=0$.

\begin{figure}[htb]
{\centering \resizebox*{1\columnwidth}{!}{\includegraphics{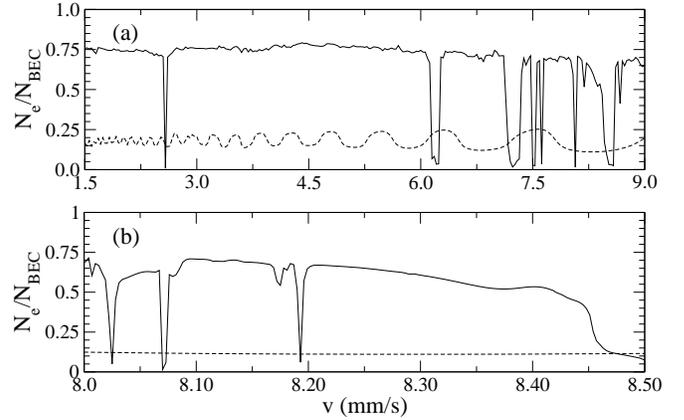}} \par}
\caption{Top: comparison between the number of extracted atoms by nonlinear (continuous 
line) and linear (dashed line) tweezers for different values of the velocity. The $y$ axis indicates 
the number of atoms trapped by the different optical tweezers normalized to the total number of atoms 
that form the initial BEC reservoir. The $x$ axis represents the velocity of the tweezers when they 
traverse the condensate. Bottom: detail from the top plot for $v=8.0$ mms$^{-1}$ to $v=8.5$ mms$^{-1}$.}
\label{fig6}
\end{figure}

\begin{figure}[htb]
{\centering \resizebox*{1\columnwidth}{!}{\includegraphics{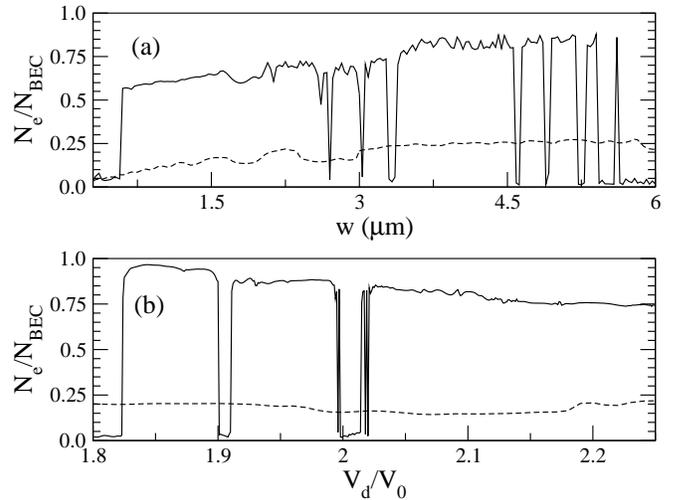}} \par}
\caption{Fraction of atoms extracted from the reservoir as a function of (a) the width $w$  and  (b) the 
intensity $I$  of the optical tweezers. The velocity was fixed to $v=5$ mm s$^{-1}$ in all cases. 
The continuous lines represent the data for the nonlinear optical tweezers. The dotted lines
correspond to the linear case.}
\label{fig7}
\end{figure}

The simulations reveal that, in the nonlinear case, the fraction of extracted atoms $N_e/N_{BEC}$ depends 
crucially on the beam parameters, making the process highly controllable by changing the velocity, 
intensity or width of the beam. In Fig. \ref{fig6} we show the dependence of the percentage of extracted atoms 
on the velocity of the beams. The black dashed line represents the values obtained with the linear 
tweezers. The continuous black line shows the dependence of the extracted atoms on the velocity in 
the nonlinear case. As it can be appreciated in the plot, the efficiency of the nonlinear optical tweezers 
is much higher than in the linear case. Another dramatic difference in 
both configurations is the presence of sharp variations of $N_e$ at some velocities in the nonlinear 
configuration, allowing more control on the number of extracted atoms. 

In Figs. \ref{fig7} we plot  the variation 
of $N_e/N_{BEC}$ versus $w$ [Fig. \ref{fig7}(a)] and the depth of the dipole optical potential $V_d$, measured 
in units of $V_0=\hbar\nu_\perp/2$ [Fig. \ref{fig7}(b)]. In both cases, the data were obtained by fixing the velocity 
of the beams to $v=5$ mm s$^{-1}$. As in Fig. \ref{fig6} the continuous lines refer to the nonlinear tweezers and 
the dashed lines to the linear ones. As it can be appreciated in the captions, the behavior is similar 
to the variations observed in Fig. \ref{fig6}. The effect of nonlinear interactions is the existence
of sharp variations in the number of extracted atoms at certain values of the width and intensity of the
laser tweezer. This adds extra control possibilities which are not accessible with linear traps. 

From the practical point of view the regions with complex variations of the number of particles as a function 
of the parameters (for example the range of potentials around $V_d/V_0\approx 2.0$ in Fig. \ref{fig7}(b))
are interesting in order to provide control on very few number of atoms. This effect opens new possibilities of achieving 
creation of macroscopic superposition of quantum states from arrays of Bose-Einstein condensates \cite{phillips}.
On the other hand, these regions are reminiscent of the fractal windows in  chaotic scattering and 
resonances in soliton collisions typical of the interactions of nonlinear waves, which have received a lot of 
attention recently \cite{Chao}.

\section{conclusions}

In summary, we have proposed a novel mechanism for extracting atoms 
from a BEC reservoir. Our system uses optical tuning of nonlinear interactions 
between atoms to extract them from the trap. 

By means of numerical simulations of the mean-field model equations, we have shown that this optically-induced 
spatial variation of the scattering length allows to control the number of atoms extracted and the position of 
the outcoupled solitary wavepacket in a simple and robust way. 

We have also compared our 
nonlinear tweezer concept  with the action of a linear potential as the one generated by ordinary laser beams  
in order to illustrate the crucial effect of nonlinear
interactions in the process. We have also described chaotic scattering of solitons.

Our results provide new ways to control the preparation of BECs with a controllable number of atoms and
are fully testable with current BEC experiments and can be easily generalized to systems with different atomic species.

\acknowledgments 

This work has been partially supported by Ministerio de Educaci\'on y Ciencia, Spain
(projects FIS2004-02466, FIS2006-04190, network
FIS2004-20188-E), Xunta de Galicia (project PGIDIT04TIC383001PR) and Junta de 
Comunidades de Castilla-La Mancha (project PAI-05-001).


\end{document}